\documentclass[aps,twocolumn,superscriptaddress]{revtex4}

\usepackage{amsmath,amssymb,graphicx}
\usepackage{color}

\definecolor{yblue}{rgb}{0.06, 0.3, 0.57}
\usepackage{hyperref}
\hypersetup{colorlinks=true,linkcolor=yblue,citecolor=yblue,urlcolor=yblue}

\graphicspath{{fig-ps/}}

\begin{document}

\title{An Adiabatic Invariant Approach to Transverse Instability: \\
  Landau Dynamics of Soliton Filaments}

\author{P. G. Kevrekidis}
\affiliation{Department of Mathematics and Statistics, University of Massachusetts,
Amherst, Massachusetts 01003-4515 USA}

\author{Wenlong Wang}
\email{wenlongcmp@gmail.com}
\affiliation{Department of Physics and Astronomy, Texas A$\&$M University,
College Station, Texas 77843-4242, USA}

\author{R. Carretero-Gonz{\'a}lez}
\affiliation{Nonlinear Dynamical Systems
Group,\footnote{\texttt{URL}: http://nlds.sdsu.edu}
Computational Sciences Research Center, and
Department of Mathematics and Statistics,
San Diego State University, San Diego, California 92182-7720, USA}

\author{D. J. Frantzeskakis}
\affiliation{Department of Physics, National and Kapodistrian University of Athens,
Panepistimiopolis, Zografos, 15784 Athens, Greece}

\begin{abstract}
  Consider a lower-dimensional solitonic structure embedded in a higher
  dimensional space, e.g., a 1D dark soliton embedded in 2D space, a
  ring dark soliton in 2D space, a spherical shell soliton in 3D space etc.
  By extending the Landau dynamics approach 
  [Phys. Rev. Lett. {\bf 93}, 240403 (2004)], we show that it is possible to capture the transverse dynamical
  modes (the ``Kelvin modes'') of the undulation of this
  ``soliton filament'' within the higher dimensional space.
  These are the transverse stability/instability modes and are the
  ones potentially responsible for the breakup of the soliton into
  structures such as vortices, vortex rings etc. We present the theory
  and case examples in 2D and 3D, corroborating the results
  by numerical stability and dynamical computations.
\end{abstract}

\maketitle


{\it Introduction.} In numerous contexts, such as atomic
physics~\cite{stringari,siambook}, nonlinear optics~\cite{kivshar},
water waves~\cite{water}, 
and others~\cite{dauxois}, soliton dynamics 
is of crucial importance. It is then especially
relevant, e.g., in 
external 
potentials 
confining Bose-Einstein condensates (BECs) or in 
refractive index landscapes in optics 
or, more recently, in gain/loss profiles of $\mathcal{PT}$-symmetric
media~\cite{yang}, to be able to characterize the reduced degree-of-freedom
evolution of soliton characteristics (center of mass, width, etc.~\cite{kivmal,malomed}).

While mathematical theories including those of nonlinear dispersive wave
equations, such as the nonlinear Schr{\"o}dinger (NLS)
equation~\cite{sulem,ablowitz}, are well developed in one-dimensional
(1D) settings, 
solitons often emerge in (or experimental settings
naturally feature) higher dimensional scenarios. Then, a question of
paramount importance is that of the stability of 
e.g., 1D or quasi-1D 
(in the case of polar or spherical
coordinates) solitonic ``filaments'' in the higher dimensional
space in which they may be embedded. A classical example of 
an instability that may arise because of the transverse degrees of freedom, is the
transverse modulational (or ``snaking'') instability, 
first analyzed in Ref.~\cite{kuzne} for dark soliton stripes embedded in a 2D space.
It has since then motivated many 
studies 
in optics 
and in atomic physics, 
exploiting as well as evading the instability,
both theoretically~\cite{smirnov,us} and experimentally~\cite{brian}.

In this work, our aim 
is to develop a theory that
combines these two elements: considers the soliton motion, 
but explores it in a scenario where the
solitonic structure is embedded in a higher dimensional space.
Relevant examples include a 1D (rectilinear) soliton in a 2D domain,
as is the case 
of Ref.~\cite{kuzne}, as well as two quasi-1D dark soliton structures: 
the ring dark soliton (RDS) and the dark spherical shell soliton (SSS), 
in 2D and 3D settings, respectively. RDSs were 
%
predicted \cite{kivyang} (and observed \cite{rings}) in optics,  
and in BECs~\cite{rings2}, while SSSs were studied in optics and BECs~\cite{kivyang,carr,wenlong},  
and observed as transient structures in a BEC experiment \cite{hau}.
In earlier works, the so-called
{\it Landau dynamics} approach (i.e., a semiclassical dynamics of the solitary
wave as a quasi-particle relying on a local-density approximation)
was developed for dark solitons~\cite{konotop_pit,konotop_pit2}
and for RDSs~\cite{korneev}.
Here, we adapt this approach towards accounting
for the possibility of transverse instabilities, i.e., for
transversely undulating soliton filaments. %
Employing this suitably modified technique, we 
analyze the (Kelvin) modes of transverse undulation,
identify the modes of potential instability, and 
also offer a previously unknown class of partial differential
equations (PDEs) unveiling the motion of the soliton filament
within the higher dimensional space. The diversity of the above
examples will serve to illustrate the broad applicability of
this concept, 
well beyond the specific selections made herein.

{\it Background and  
Transverse Instability of
  Planar Dark Solitons.}
%
Our starting point is the dimensionless (for relevant adimensionalizations
in the BEC problem, see Refs.~\cite{stringari,siambook}) form of
the 1D defocusing NLS equation:
\begin{eqnarray}
  i u_t =-\frac{1}{2} u_{xx} + |u|^2 u + V(x)\, u.
  \label{dark1}
\end{eqnarray}
In the absence of the external potential ($V \equiv 0$), and assuming 
a background density $\mu$, the dark soliton preserves the renormalized energy 
(see, e.g., the review~\cite{djf}):
%
\begin{eqnarray}
  H_{\rm 1D}=\frac{1}{2} \int_{-\infty}^{\infty} \left[ |u_x|^2 +
  \left(|u|^2-\mu \right)^2 \right] dx.
  \label{dark2}
\end{eqnarray}
For the dark soliton solution of Eq.~(\ref{dark1}) (for $V=0$), with center 
$x_0$ and velocity $v=\dot{x}_0$, namely:  
\begin{eqnarray}
u=e^{-i \mu t} \left[
  \sqrt{\mu - v^2} \tanh \left(\sqrt{\mu - v^2} (x-x_0) \right) + i v \right],~
\label{dark2a}
\end{eqnarray}
the energy reads 
$H_{\rm 1D}=(4/3) (\mu-\dot{x}_0^2)^{3/2}$. Then, 
according to the Landau dynamics approach~\cite{konotop_pit,konotop_pit2} 
this energy is treated as an {\it adiabatic invariant} in the presence
of a slowly varying potential, i.e., 
the background density
$\mu$ will be slowly varying according to $\mu \rightarrow \mu - V(x)$. 
Then, assuming the adiabatic invariance of
\begin{eqnarray}
  H_{\rm 1D} = \frac{4}{3} \left(\mu - V(x_0) -\dot{x}_0^2 \right)^{3/2} ~\Rightarrow~
  \ddot{x}_0=- \frac{1}{2} V'(x_0),
  \label{dark3}
  \end{eqnarray}
i.e., by its direct differentiation, one obtains the effective equation
for the dark soliton center, 
in remarkable  
agreement with numerical results~\cite{konotop_pit,konotop_pit2,siambook,djf}.

Our proposal is to  extend this notion of adiabatic invariants
to the case of 1D stripes 
and quasi-1D RDSs
embedded in 2D, as well as SSSs 
embedded in 3D.
Our ultimate aim is not only to provide a fresh and broad/general perspective
on the transverse instability, but also to describe the motion of
the soliton filament in the 
higher dimensional space. It is known that in 2D 
as the undulation intensifies along the transverse direction, 
the soliton eventually decays into vortex-antivortex pairs~\cite{smirnov,dep}.
%
In 3D, the corresponding transverse instability gives rise to vortex
rings and vortex lines~\cite{siambook,brian,pindzola,wenlong}.

We specifically propose to consider
the energy of a dark soliton stripe/filament again as an
adiabatic invariant, whereby the 2D energy has an
additional term:
%
\begin{eqnarray}
  H_{\rm 2D}=\frac{1}{2}  \int_{-\infty}^{\infty} \left[ |u_x|^2 + |u_y|^2 +
  \left(|u|^2-\mu \right)^2 \right]dx dy.
  \label{dark4}
\end{eqnarray}
Now, assuming an ansatz of the form of Eq.~(\ref{dark2a}) with
the center position not solely a function of $t$, but
a function $x_0(y,t)$, we obtain an ``effective energy'' (an adiabatic
invariant, again) of the form:
 \begin{eqnarray}
  E= \frac{4}{3} \int_{-\infty}^{\infty} \left(1 + \frac{{x_0}_y^2}{2} \right)
  \left(\mu - V(x_0) -{x_0}_t^2 \right)^{3/2} dy.
  \label{dark5}
\end{eqnarray}

Based on this ``effective Hamiltonian'', one can describe the {transverse
motion of the soliton filament}. At the level of existence and
stability of equilibria, one can use $x_0(y,t)=X_0(t) + \varepsilon
X_1(t) \cos(n y)$, in order to (a) identify the leading order
dynamical equation above [cf.~second of Eqs.~(\ref{dark3})], 
and (b) obtain the small amplitude ---longitudinal, as well as transverse---
excitations 
\footnote{%
The small amplitude excitations are described at the level of the Hessian, 
i.e., at O$(\varepsilon^2)$, by the energy per unit length (in
each periodic cell of size $L$) expansion:
\\
$
\frac{E}{L}=E_0+  \varepsilon^2 
\left\{
\frac{1}{2} \dot{X}_1^2
 + \frac{1}{2} X_1^2
\right.
$
\\
\null
\qquad
$
\left.
\times
  \left[-\frac{1}{3} (\mu - V(X_0))\, n^2
-\frac{1}{2} \frac{V'(X_0)^2}{\mu - V(X_0)} 
    + \frac{1}{2} V''(X_0) \right]\right\},
$
\\
where $E_0=(\mu-V(X_0)-{X_0}_t^2)^{3/2}$. This suggests that the term in 
the bracket, evaluated at $V'(X_0)=0$ for equilibrium points,
constitutes the squared frequency of such eigenmodes. 
}. 
Perhaps even more importantly,
one can obtain from energy conservation
the field-theoretic equation governing such a modulation.
To simplify the exposition, for $V=0$, 
the relevant equation reads (see Supplement for details):
\begin{eqnarray}
  \left(1 + 
\frac{{x_0}_y^2}{2} \right)
     {x_0}_{tt} = \frac{1}{3} {x_0}_{yy} \left( \mu - {x_0}_t \right)
     + {x_0}_t {x_0}_y {x_0}_{ty}.
  \label{dark7}
\end{eqnarray}
Importantly, at the {\it linear} level,
this PDE is elliptic [${x_0}_{tt} + (\mu/3) {x_0}_{yy}=0$]
leading to the instability rate 
of Ref.~\cite{kuzne}. However, the key feature is that this novel,
nonlinear PDE, 
Eq.~(\ref{dark7}), and its variant in the presence
of the trap,  
can describe the nonlinear evolution of the
soliton filament. When the latter
leads to a jump discontinuity (a shock), then the
filament breaks and develops the vortical patterns of principal
interest herein.



\begin{figure}
\includegraphics[height=4.5cm]{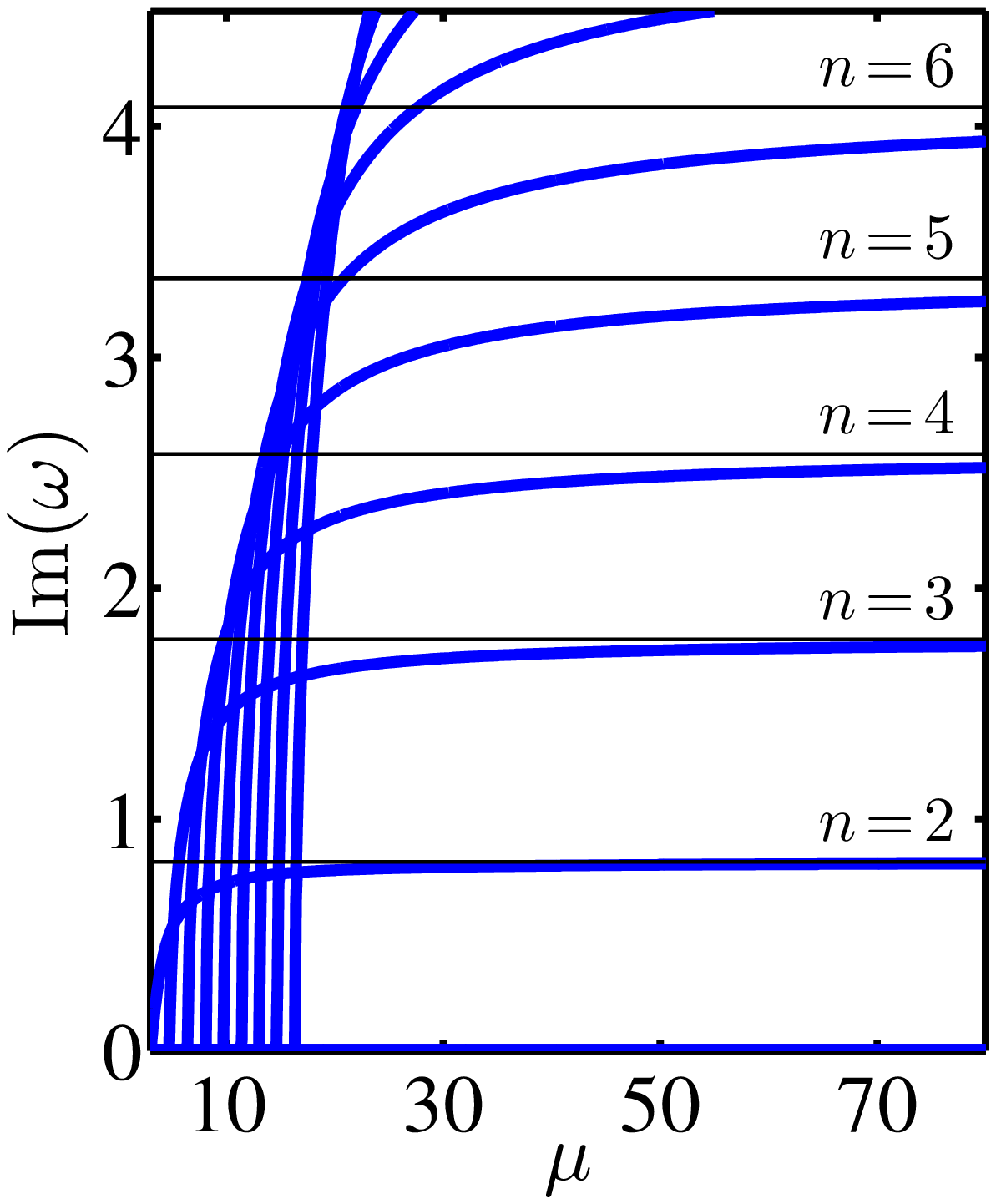}
\includegraphics[height=4.5cm]{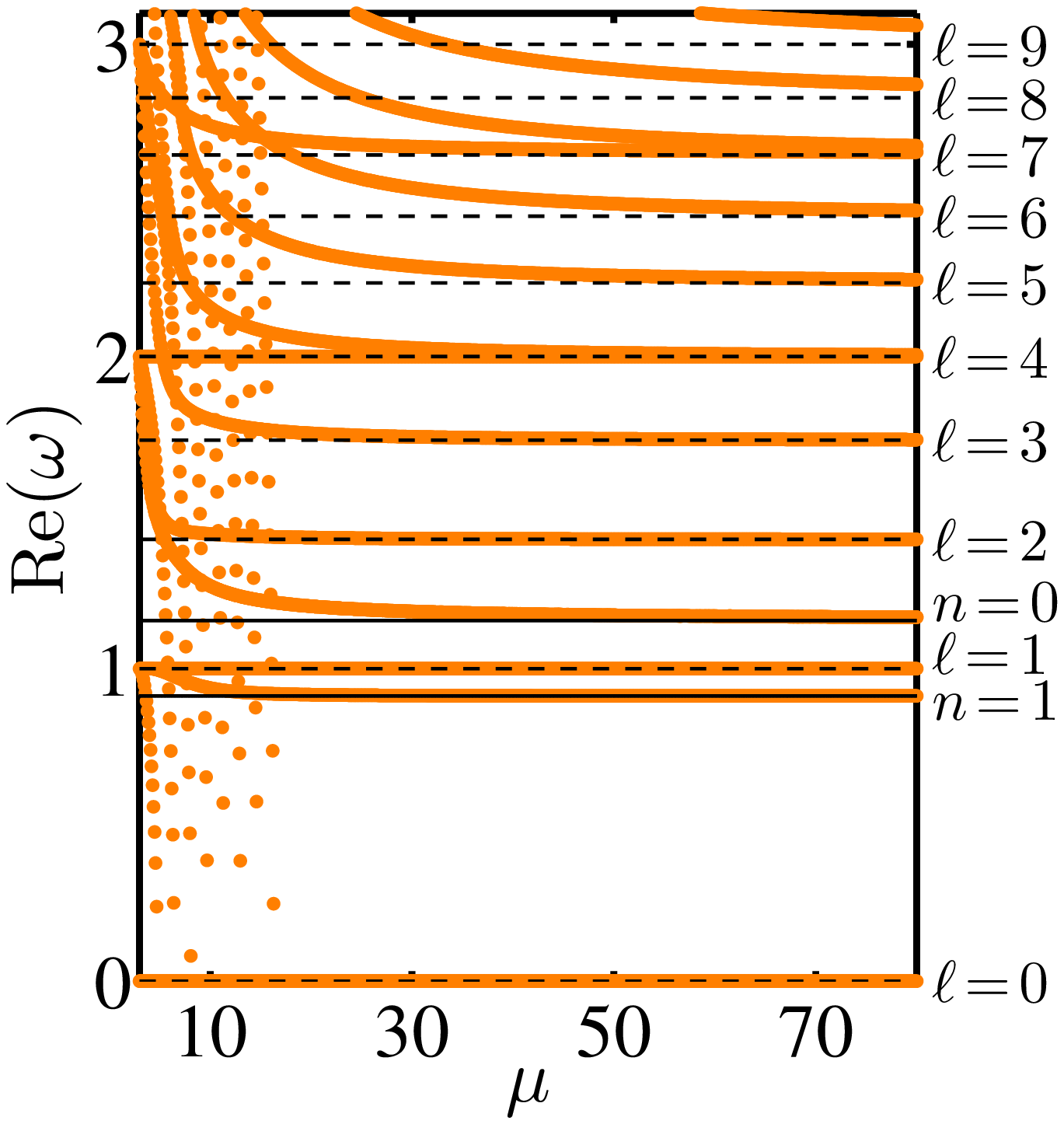}
\caption{(Color online) 
Imaginary (left) and real (right) parts of the BdG 
excitation spectrum for the ring dark soliton (RDS).
The former correspond to the instability modes, while the latter to 
the vibration modes.
The thin horizontal solid lines in both panels correspond to
undulation modes of the RDS in the Thomas-Fermi (large nonlinearity/large 
chemical potential $\mu$) limit, as predicted by Eq.~(\ref{DSR_predict}).
The thin horizontal dashed lines in the right panel correspond to
the asymptotic predictions for the ground state spectrum for $k=0$
and $0\leq\ell\leq 9$ [cf.~Eq.~(\ref{ei2d})]. To avoid clogging the
relevant diagram, only the lowest modes of instability are shown.
%
}
\label{fig:DSR}
\end{figure}

{\it A 2D Scenario:
the Ring Dark Soliton (RDS).}
The RDS~\cite{kivyang,rings,rings2,korneev} is 
a quasi-1D solitonic structure (localized along the radial direction and extending as a filament
along the azimuthal direction) 
embedded in 2D space; 
namely, a RDS is a circular dark soliton that closes into itself.
%
%
The work of Ref.~\cite{korneev} utilized 
the argument of Ref.~\cite{konotop_pit} 
to a RDS of radius 
$R(t)$ and used 
the adiabatic invariant $E=2 \pi R (\mu -\dot{R}^2 - V(R))^{3/2}$ 
to obtain its {purely radial} equation of motion
(see Supplement for details).
Our Landau dynamics 
generalization considers this as a genuinely 2D
filament whose radial position is 
$R(\theta,t)$, i.e., includes azimuthal
``departures'' from a perfect ring. Then, from the
polar form of Eq.~(\ref{dark4}), the adiabatic
invariant is generalized as:
\begin{eqnarray}
  E=\int_0^{2 \pi} R \left(1 + \frac{R_{\theta}^2}{2  R^2} \right)
  \left(\mu -R_t^2 - V(R) \right)^{3/2} d \theta.
  \label{dark8}
\end{eqnarray}
From this 
equation, we can extract conditions for the
existence of stationary RDS
filaments of radius $R_0$ in a certain potential,
and the stability (eigenfrequencies $\omega$)
of small-amplitude excitations around it, using
$R=R_0 + \varepsilon\, e^{i (n \theta+\omega t)}$. These, respectively,
read:
\begin{eqnarray}
  3 R_0 V'(R_0)  &=& 2 (\mu - V(R_0)),
  \label{equil}
  \\[2.0ex]
  \omega^2 &=&  \frac{V'(R_0)}{2 R_0} \left[\frac{5}{3} -n^2 + \frac{R_0 V''(R_0)}{V'(R_0)} \right].~~~
  \label{dark9}
\end{eqnarray}
For the experimentally generic in BECs case of a parabolic
potential, $V(R)=(1/2) \Omega^2 R^2$~\cite{stringari,siambook}, we obtain
\begin{equation}
\displaystyle
R_0^2=\frac{\mu}{2\, \Omega^2}
\quad{\rm and}\quad 
\omega= \pm \left(\frac{1}{2}\left(\frac{8}{3}-n^2\right)\right)^{1/2} \Omega.
\label{DSR_predict}
\end{equation}
in {\it very good}
agreement with asymptotic predictions of our numerical computations,
as shown in Fig.~\ref{fig:DSR} (see also for the equilibrium
radius Fig.~14 in Ref.~\cite{kaper}).
It is important to highlight here that the above
prediction enables a systematic and {\it complete} understanding
of the modes of the 
Bogolyubov-de Gennes (BdG) linearization
analysis in the asymptotic limit where this particle description
is relevant, namely the 
Thomas-Fermi (TF) limit~\cite{stringari,siambook}.
This is due to the following fact:
the spectrum of a nonlinear wave consists of the spectrum of the
underlying ``background'' (the fundamental, equilibrium state on top
of which the excitation exists) and the localized point spectrum
associated with the excitation (the internal undulations
of the solitary wave). In the BEC realm, the spectrum
of the underlying ground state has been revealed in the
fundamental work of Ref.~\cite{stringari2} (see also Ref.~\cite{dep2})
and in 2D consists of the eigenfrequencies:
\begin{equation}
\omega=\pm \Omega (\ell + 2k (1+\ell) + 2 k^2)^{1/2}, 
\label{ei2d}
\end{equation}
for $k,\ell \geq 0$ 
(thin horizontal dashed lines in Fig.~\ref{fig:DSR}).
Hence, the union of this set and of the eigenfrequencies
of Eq.~(\ref{DSR_predict}) 
(thin horizontal solid lines in Fig.~\ref{fig:DSR}) 
provides an unprecedented, {\it all-encompassing}
theoretical prediction for the BdG spectrum 
of a RDS in the TF limit.


\begin{figure*}[thb]
\includegraphics[width=17.55cm]{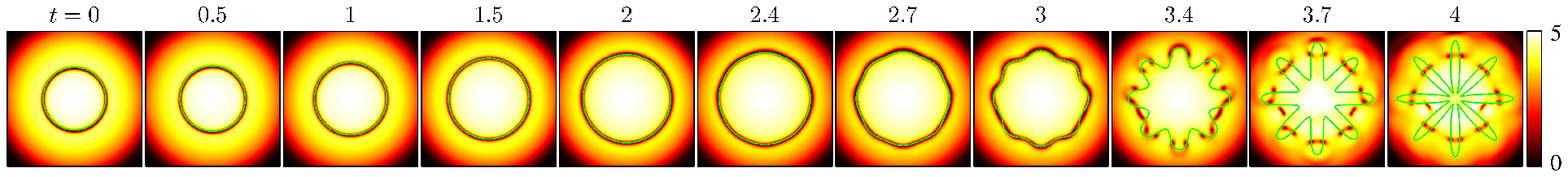}\\[-0.4ex]
\includegraphics[width=17.55cm]{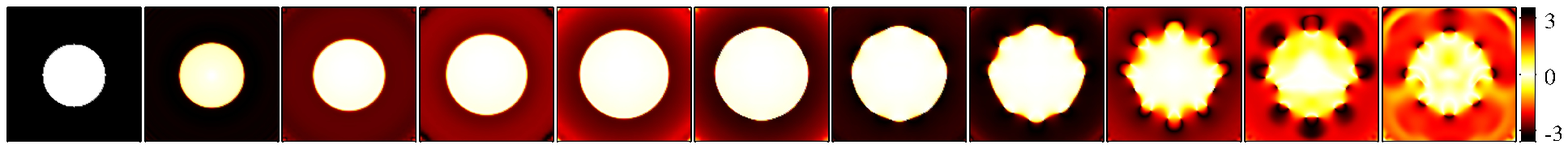}\\[ 1.0ex]
\includegraphics[width=17.55cm]{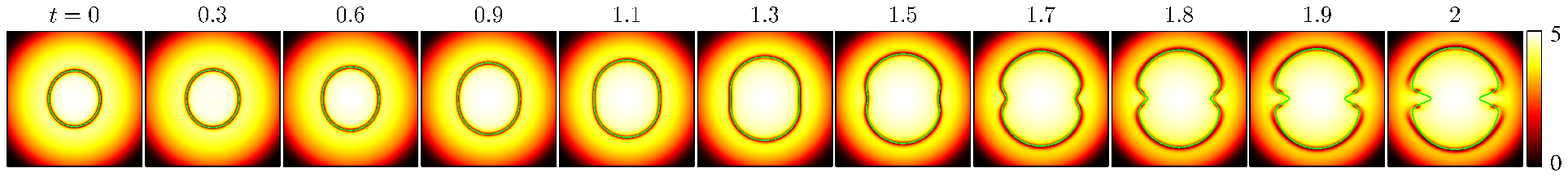}   \\[-0.4ex]
\includegraphics[width=17.55cm]{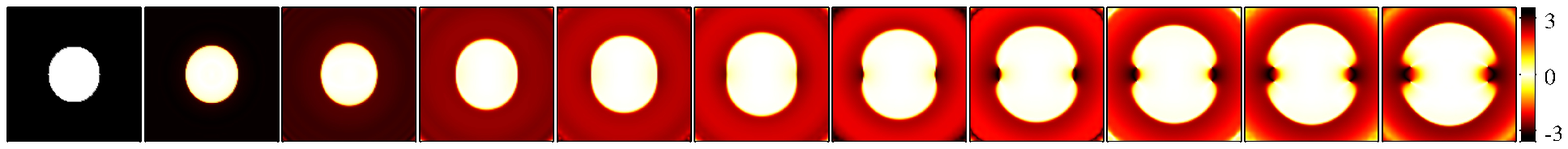}
\caption{(Color online) 
Comparison of the full NLS dynamics (\ref{dark1}) and the
adiabatic invariant PDE obtained from Hamilton's principle
applied to Eq.~(\ref{dark8}) for a ring dark soliton (RDS) for $\mu=24$.
Top series of panels: RDS with initial azimuthal position 
given by $R(\theta)=2.8+0.1\cos(\theta)+3\times 10^{-7}\cos(8\theta)$.
Bottom  series of panels: another RDS, but now with
initial azimuthal position given by $R(\theta)=2.4+0.1\cos(2\theta)$.
In both series of panels the top (bottom) subpanels depict the 
density (phase) of the solution on a $6\times 6$ square.
The dynamics of the ensuing PDE is depicted by a solid green line.
Note how the derived PDE closely follows the 
RDS dynamics before the break up of the latter
into a series of vortex pairs.
}
\label{Dynamics}
\end{figure*}

Going one step beyond the equilibrium radius and the
near-equilibrium vibrations, one can study the ring
PDE evolution for $R(\theta,t)$. This can be obtained from
energy conservation
applied to Eq.~(\ref{dark8}). To illustrate this, 
we 
present two examples of dynamics 
comparing Eqs.~(\ref{dark1}) with the dynamics resulting
from the energy functional (\ref{dark8}), as shown in Fig.~\ref{Dynamics}. 
Both cases correspond to non-ideal rings in the TF limit, 
for $\mu=24$ and $\Omega=1$. 
In one case, we have perturbed the rings with a combination 
of $n=0$, $n=1$, and $n=8$ modes with the initial condition 
$R(\theta)=2.8+0.1\cos(\theta)+3\times 10^{-7}\cos(8\theta)$ 
(top series of panels). 
In the second case, the ring is perturbed using a combination 
of $n=0$ and $n=2$ modes with the  initial condition 
$R(\theta)=2.4+0.1\cos(2\theta)$ (bottom series of panels). 
The dynamics compare extremely well for the two PDEs until the rings 
significantly deform/break, illustrating that the PDE derived from
the energy functional (\ref{dark8}) is a valuable tool for understanding
the evolution of ring soliton filaments. 
%
%
See Ref.~\cite{movies} for movies
depicting the dynamics of these two examples and
more definitively showcasing the comparison.

{\it A 3D 
Scenario: Dark Spherical Shell Soliton (SSS).} 
Extending our considerations 
to a 3D case, we first study 
a 1D (rectilinear) dark soliton 
embedded (as a planar filament) in 3D space. For the 
soliton of Eq.~(\ref{dark2a}) with center $x_0=x_0(y,z)$, the relevant
energy functional reads:
\begin{eqnarray}
  E\!\!=\!\!\int\!\! \left(1 + \frac{{x_0}_y^2}{2}+ \frac{{x_0}_z^2}{2}\right)\!
  \left(\mu - V(x_0) -{x_0}_t^2 \right)^{3/2}\!  dy dz.~~
  \label{dark3D1}
\end{eqnarray}
However, we do not focus on this simpler case (somewhat
analogous to the 2D one), but rather consider the more
intricate example of a quasi-1D dark SSS~\cite{wenlong}, 
localized along the radial
direction with a center potentially undulating according
to $R=R(\theta,\phi)$. Here, the adiabatic invariant of Landau
dynamics subject to transverse undulations assumes the form:
\begin{eqnarray}
  E=\int R^2 \left(1 + \frac{R_{\theta}^2}{2  R^2} +
  \frac{R_{\phi}^2}{2  R^2 \sin^2(\theta)} \right)
\nonumber
\\
  \times\left(\mu -R_t^2 - V(R) \right)^{3/2} d \theta d \phi.
  \label{dark3D2}
\end{eqnarray}
While the radial dynamics can be straightforwardly obtained~\cite{wenlong},
the calculation of
the internal undulation modes is 
more technically involved,
incorporating a decomposition of $R=R_0 + \varepsilon R_1(\theta)
R_2(t) \cos(n \phi)$ effectively into spherical harmonic modes.
The shell's equilibrium position $R_0$ 
satisfies:
\begin{eqnarray}
  3 R_0 V'(R_0)= 4 (\mu - V(R_0)).
  \label{dark3D3}
\end{eqnarray}
%
For the physically relevant (isotropic) parabolic potential 
$V(R)=(1/2) \Omega^2 R^2$, this result leads to the equilibrium
position $R_0=(4 \mu/(5 \Omega^2))^{1/2}$ in very
good agreement with numerical results~\cite{wenlong}. The
distilled final expression of the far more tedious eigenmodes
calculation reads:
\begin{eqnarray}
  \frac{\omega^2}{\Omega^2}= \frac{7}{6} \frac{V'}{R_0} + \frac{V''}{2}
  -\frac{V'}{4R_0} \left(\frac{B}{A} + n^2 \frac{C}{A}\right),
  \label{dark3D4}
\end{eqnarray}
where $A=\int_0^\pi R_1^2 \sin \theta d\theta$,
$B=\int_0^\pi (R_1')^2 \sin \theta d\theta$ and
$C=\int_0^\pi R_1^2 \sin \theta d \theta$, 
while $V'$ and $V''$ are evaluated at $R_0$.
In this expression $R_1=P_n^{l}(\cos(\theta))$ corresponds to the associated
Legendre polynomials. A closed form expression can
be obtained, e.g., for $A=2 (l+n)!/((2 l+1) (l-n)!)$, yet generally
these expressions amount to straightforward integral evaluations.
In the particular case of the (spherical) parabolic trap, the
expression becomes
\begin{eqnarray}
  \omega^2 = \Omega^2 \left(\frac{5}{3}- \frac{1}{4} \left(\frac{B}{A} + n^2
  \frac{C}{A}\right)\right).
  \label{dark3D5}
\end{eqnarray}

\begin{figure}
\includegraphics[height=4.5cm]{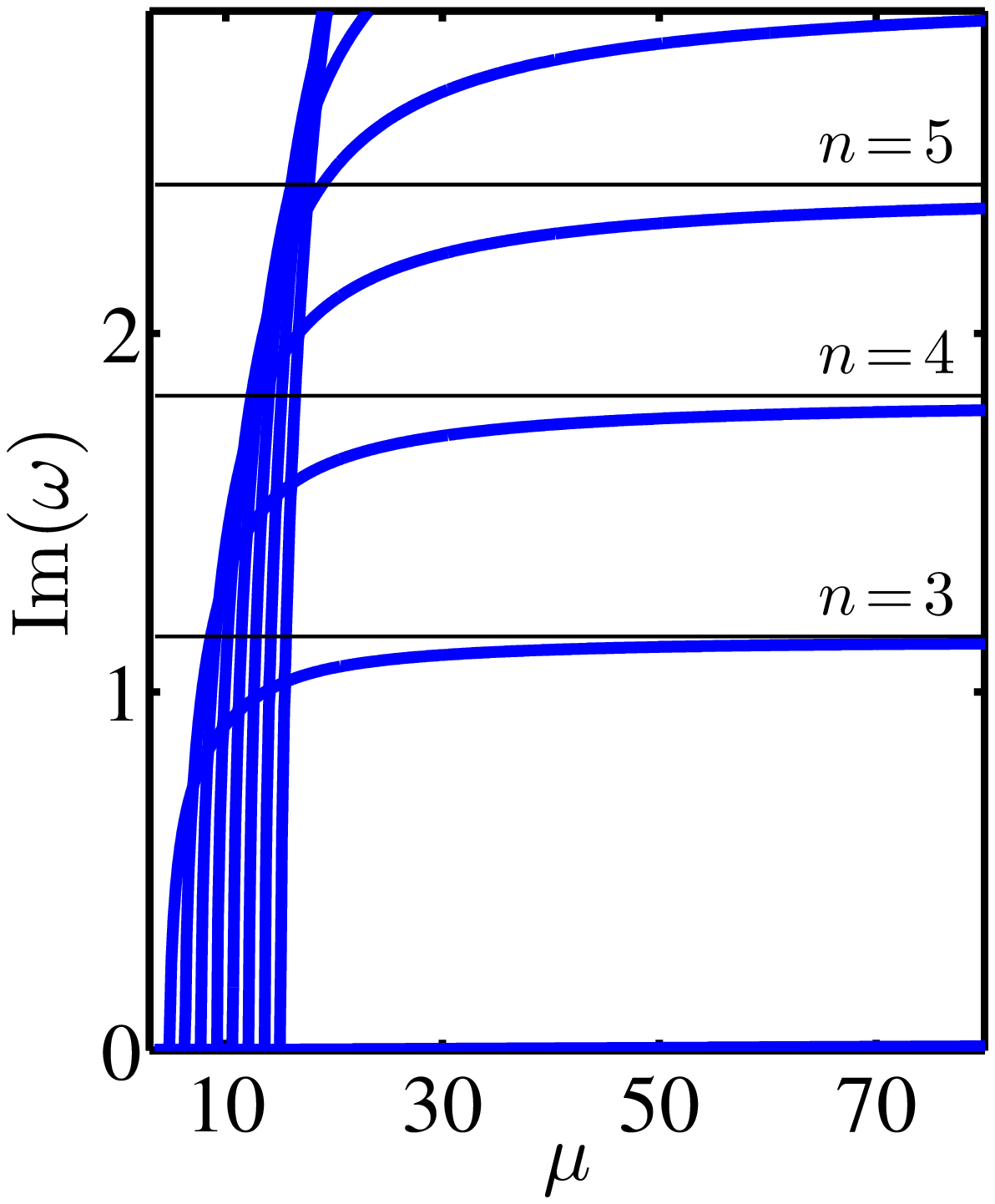}
\includegraphics[height=4.5cm]{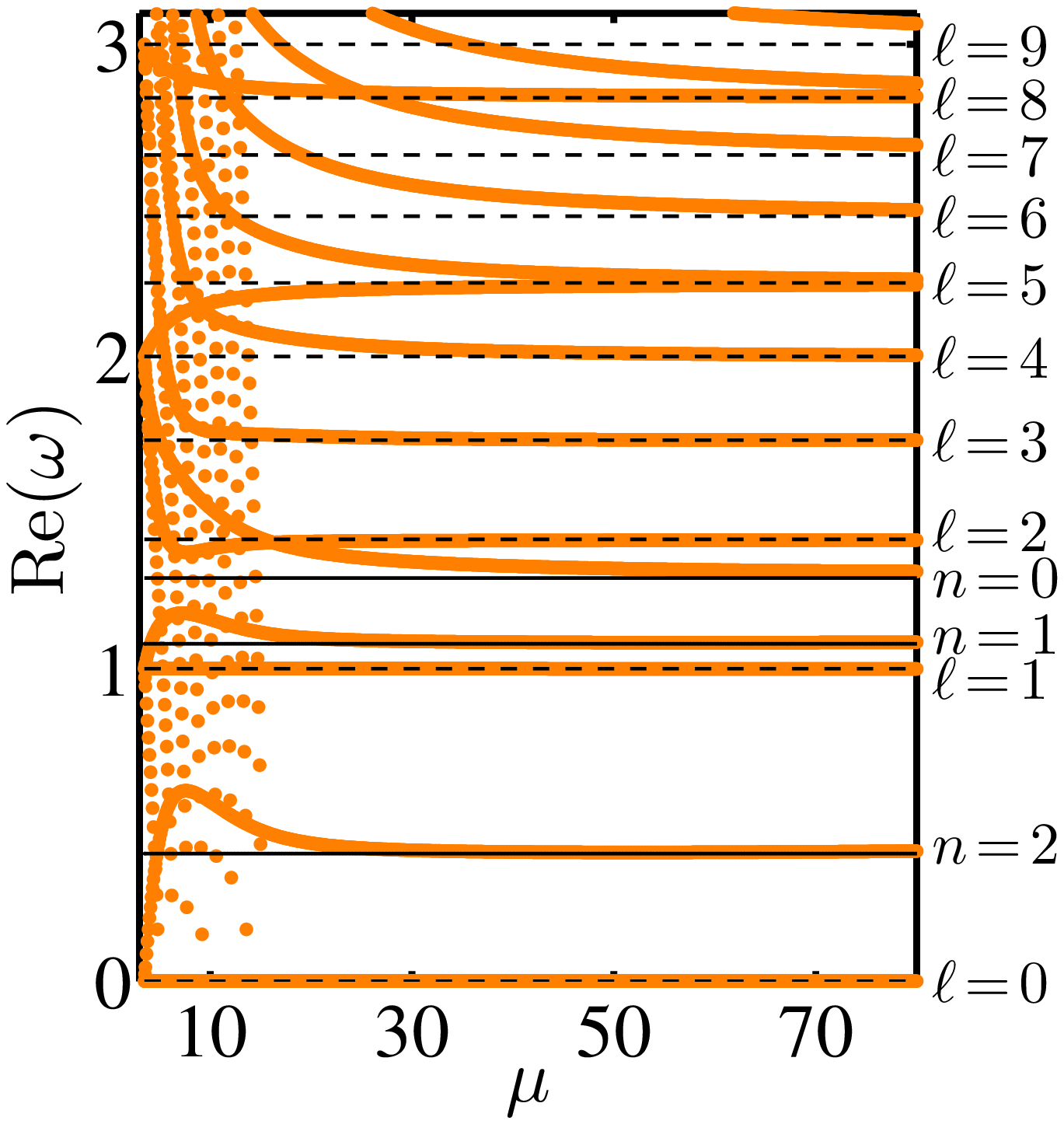}
\caption{(Color online) 
The BdG 
excitation spectrum for the dark spherical shell soliton. 
The 
layout and meaning is the same as in the case of the ring dark 
soliton of Fig.~\ref{fig:DSR}.
The thin horizontal solid
lines correspond to the undulation modes in the TF limit, 
as predicted by Eq.~(\ref{dark3D5}), and the thin
horizontal dashed lines to the asymptotic prediction for the
ground state for $k=0$ and $0\leq\ell\leq 9$ [cf.~Eq.~(\ref{ei3d})].
Once again, only the dominant instability modes are shown.
}
\label{fig:SSS}
\end{figure}

In this case too, combining the predictions of Eq.~(\ref{dark3D5})
with those for the vibration modes of the background (the ground
state of the system), namely~\cite{stringari2,dep2}:
\begin{equation}
\omega=\pm \Omega (\ell + 3 k + 2 k \ell + 2 k^2)^{1/2},
\label{ei3d}
\end{equation}
with $k, \ell \geq 0$, one obtains the {\it full} linearization (BdG) spectrum of a dark
SSS in the Thomas-Fermi limit. The relevant comparison
illustrating once again the very good asymptotic agreement
can be found in Fig.~\ref{fig:SSS}.

{\it Conclusions $\&$ Future Work.} In the present study, we provided
a generalization of the theory of Landau dynamics as applied
to solitary waves. We illustrated that this methodology can be
extended to the formulation of soliton filaments and has the ability
to provide a wealth of systematic results: (a) it can provide
quantitative information about the equilibrium features of
the solitonic filaments; (b) it can systematically characterize
their stability in the form of undulation (Kelvin) modes,
unveiling an unprecedented and complementary perspective
on the transverse instability features; 
and (c) 
it can elucidate the dynamics proper of the solitonic filament,
yielding insights on the potential amplification and eventual
breakdown (i.e., formation of vortices and vortex rings) of 
the soliton filaments.
Importantly, it should be underlined that our aim is {\it not}
to manifest the advantages of this method
over other methods (e.g., soliton perturbation theory~\cite{kivmal}
or the variational approach~\cite{malomed}) that treat solitary waves
as quasi-particles. It is rather to highlight that some of these
approaches (in fact, the variational approach can be similarly
extended to address transverse instabilities and soliton filament
motion) can be suitably ``embedded'' in higher dimensions and thus enable
the detailed characterization of soliton filament motion.

This emerging theory of soliton filaments is promising in a
broad range of directions. In addition to its usefulness
for studying the modes of droplets and dynamics in other
(e.g., magnetic~\cite{hoefer}) systems, it can be used
to obtain further insights on the dynamics of vortex
and vortex-ring creation. From the point of view of
the energy functionals and resulting PDEs for the solitonic
filaments, the emergence of these other coherent structures
is nothing but a {\it collapse} phenomenon. Hence, studying
(possibly self-similar or asymptotically
self-similar) collapse~\cite{sulem,galaktionov}
of the resulting PDEs is a particularly appealing
future topic. Moreover, the technique is by no means constrained
to dark solitons; for instance, as was discussed in Ref.~\cite{JacksonMcCannAdams},
using the energy of a vortex and a suitable azimuthal integration,
one can obtain the energy of a vortex ring. If now the center of
the vortex (in $R,Z$) varies as $R=R(\theta)$ and $Z=Z(\theta)$, then
the undulations of a vortex ring (Kelvin modes)
can be identified and compared
with the corresponding BdG results~\cite{horng,bisset}.

As an aside, we should mention here that the work of Ref.~\cite{smirnov}
presents an alternative to the present method that may be especially
powerful due to its intrinsic character. Considering the soliton filament
as a curve if embedded in 2D (or as a surface if embedded in 3D), one
can derive effective reduced PDEs for intrinsic quantities such as the
curvature (or the torsion), as well as its coupling to the speed of
the motion of the curve (or surface). This is a particularly appealing
formulation to pursue in the near future.

Finally, it would be relevant to consider extending the adiabatic invariant
approach to include the effects of dissipation. Such a generalization would
need to involve a Hamiltonian formulation for PDEs similar to the Lagrangian
formulation used in Ref.~\cite{Panos_NCVA} for $\mathcal{PT}$-symmetric
sine-Gordon and $\phi^4$ models including gain and loss and in 
Ref.~\cite{Julia_NCVA} for NLS models including gain and loss, 
which in turn, were inspired by the work of Ref.~\cite{Galley}
on classical, finite dimensional, mechanics of nonconservative systems;
see also~\cite{galley2}. 
Some of these directions are presently under consideration
and will be reported in future works.

{\it Acknowledgments.}
P.G.K.~gratefully acknowledges support
from the Alexander von Humboldt Foundation,
the US-NSF under grants DMS-1312856, and PHY-1602994.
%
W.W. acknowledges support from NSF-DMR-1151387. The work of W.W. is supported in part by the Office of the Director of National Intelligence (ODNI), Intelligence Advanced Research Projects Activity (IARPA), via MIT Lincoln Laboratory Air Force Contract No.~FA8721-05-C-0002. The views and conclusions contained herein are those of the authors and should not be interpreted as necessarily representing the official policies or endorsements, either expressed or implied, of ODNI, IARPA, or the U.S.~Government. The U.S.~Government is authorized to reproduce and distribute reprints for Governmental purpose notwithstanding any copyright annotation thereon.
R.C.G.~gratefully acknowledges support from
US-NSF under grants DMS-1309035 and PHY-1603058.
We thank Texas A\&M University for access to their Ada cluster.
P.G.K.~and D.J.F.~gratefully acknowledge the Stavros
Niarchos Foundation via the Greek Diaspora Fellowship Program.

\bigskip
\bigskip

\begin{center}
{\bf APPENDIX}
\end{center}

\section{Derivation of Equations for the Dark Soliton Stripe.}

Firstly, we provide here a proof of the dynamics for a
dark soliton stripe [Eq.~(7) in the main text]
starting from the corresponding adiabatic invariant
Hamiltonian. The conservation
of this Hamiltonian leads to
\begin{eqnarray}
\frac{dE}{dt}&=& \frac{4}{3} \int_{-\infty}^{\infty}
    \frac{\partial}{\partial t} \left[
      \left( 1 + \frac{{x_0}_y^2}{2} \right) \left( \mu - V(x_0) - {x_0}_t^2
      \right)^{3/2} \right] dy
\nonumber
\\
&=&0.
    \label{add_eq1}
\end{eqnarray}
Using the notation 
\begin{eqnarray}
A&\equiv& \mu - V(x_0) - {x_0}_t^2,
\nonumber
\\
\nonumber
B&\equiv& 1 + \frac{{x_0}_y^2}{2},
\end{eqnarray}
the integrand $I$ of the above integral reads
\begin{eqnarray}
  I= {x_0}_y {x_0}_{yt} A^{3/2} + \frac{3}{2} A^{1/2} B \left(-V'(x_0) {x_0}_t
  - 2 {x_0}_t {x_0}_{tt} \right).
\nonumber
\end{eqnarray}
Now, if we integrate by parts the first term in the integral, assuming
boundary conditions such that surface terms disappear, we obtain
\begin{eqnarray}
  \int_{-\infty}^{\infty} {x_0}_y {x_0}_{yt} A^{3/2} dy=-
  \int_{-\infty}^{\infty} {x_0}_t \left( {x_0}_y A^{3/2} \right)_y dy.
\nonumber
\end{eqnarray}
Now all the terms in the integral multiply ${x_0}_t$ and hence
in order for the integral to vanish for arbitrary choices of
$x_0$, the term multiplying
${x_0}_t$ should vanish. In fact, this is rather intuitive because
the reverse process (i.e., multiplying the resulting equation
by  ${x_0}_t$ and integrating over $y$) leads naturally
to the conservation law for the energy. This way, we obtain the final
PDE for the filament's dynamical evolution:
\begin{eqnarray}
  {x_0}_{tt} B \!+\! \frac{1}{3} {x_0}_{yy} A \!=\! {x_0}_y {x_0}_t {x_0}_{yt}
  \!-\!\frac{1}{2} V'(x_0) \left( B\! -\! {x_0}_y^2 \right).
 \label{add_eq4}
\end{eqnarray}
Equation~(\ref{add_eq4}) has a number of meaningful limits:
\begin{itemize}
\item Near the linear limit, as indicated in the main text,
  it leads to %
\begin{eqnarray}
{x_0}_{tt}  + \frac{\mu}{3} {x_0}_{yy} =0,
\end{eqnarray}
yielding the proper linear growth rate of the transverse instability~\cite{SSSkuzne}.
\item If $x_0=x_0(t)$ only, it recovers the celebrated
  result for the single dark soliton (obtained originally
  by Busch and Anglin~\cite{SSSbusch} and also by Konotop and
  Pitaevskii~\cite{SSSkonotop_pit}),
\item In the case where there is no potential, i.e., for $V(x_0)=0$, it yields
  Eq.~(7) in the main text, namely:
  \begin{eqnarray}
    {x_0}_{tt} B + \frac{1}{3} {x_0}_{yy} ( \mu - {x_0}_t^2)
    = {x_0}_y {x_0}_t {x_0}_{yt}.    
    \label{add_eq5}
  \end{eqnarray}
\end{itemize}

\section{Derivation of Equations for the Ring Dark Soliton.}

We proceed in a similar fashion for the derivation of the
equation of motion in the case of the ring dark soliton.
Using the notation
\begin{eqnarray}
C&\equiv& \mu - V(R) - {R}_t^2, 
\nonumber
\\
\nonumber
D&\equiv& 1+ \frac{R_{\theta}^2}{2 R^2},
\end{eqnarray}
we obtain that:
\begin{eqnarray}
  \frac{3}{4} \frac{d E}{d t}&=&
  \int_{0}^{2 \pi} \left[ R_t D C^{3/2} + \frac{R_{\theta} R_{\theta t}}{R}
  C^{3/2} - \frac{R_{\theta}^2}{R^2} R_t C^{3/2}
\right.
\nonumber
\\
&+&
\left.
  \frac{3}{2} R D C^{1/2}
  (-V'(R) - 2 R_{tt}) R_t \right] d \theta = 0.
  \label{add_eq6}
\end{eqnarray}
Following a similar pattern as the derivation in the previous
section, we recognize that all other terms multiply $R_t$ except for
the second one, for which we use integration by parts as follows:
\begin{eqnarray}
  \int_{0}^{2 \pi} \frac{R_{\theta} R_{\theta t}}{R}
  C^{3/2}  d \theta = - \int_{0}^{2 \pi} R_t
  \frac{\partial}{\partial \theta} \left( \frac{R_{\theta}}{R} C^{3/2} \right)
  d \theta.
\nonumber
\end{eqnarray}
Here, the surface terms disappear due to the periodic boundary conditions
in the angular variable. As a result, once again all terms multiply
$R_t$ and hence this multiplicative prefactor of $R_t$ in the angular
integral has to vanish for the integral to generically vanish.
The resulting PDE for the ring evolution reads:
\begin{eqnarray}
  C D - \frac{R_{\theta \theta}}{R} C &=&
  -\frac{R_{\theta}}{R} \left( \frac{3}{2} V'(R) R_{\theta} + 3 R_t
  R_{t \theta} \right) 
\nonumber
\\
&+&
 R D \left(\frac{3}{2} V'(R) + 3 R_{tt} \right)
  \label{add_eq8}
\end{eqnarray}
Again a series of special limits are of particular value:
\begin{itemize}
\item For the homogeneous steady state (time- and angle-independent
  $\theta$) $R_0$, we obtain $\mu-V(R_0)=\frac{3}{2} R_0 V'(R_0)$, in line,
  e.g., with Ref.~\cite{SSSkaper}.
\item In the case of angle-independent radial motion, we obtain
  \begin{eqnarray}
    R_{tt} = \frac{C}{3 R} - \frac{1}{2} V'(R),
    \label{add_eq9}
  \end{eqnarray}
  in agreement with Ref.~\cite{SSSkorneev}.
\item Finally, assuming $R=R_0 + \epsilon R_1(t) \cos(n \theta)$,
  we obtain to O$(\epsilon)$ the evolution for $R_1$ in the form:
  \begin{eqnarray}
    {R_1}_{tt} = \frac{V'(R_0)}{3 R_0}
    \left[ -\frac{5}{2} + \frac{3}{2} n^2 - \frac{3}{2} \frac{R_0 V''(R_0)}{V'(R_0)} \right] R_1.
    \label{add_eq10}
  \end{eqnarray}
  Assuming $R_1 \sim e^{i \omega t}$, we obtain the modes of vibration
  \begin{eqnarray}
    \omega^2 = \frac{V'(R_0)}{2 R_0}
    \left[ \frac{5}{3} -n^2 + \frac{R_0 V''(R_0)}{V'(R_0)} \right]
    \label{add_eq11}
  \end{eqnarray}
  In the case of a parabolic trap, $V(r)=\frac{1}{2}\Omega^2 r^2$,
  this leads  to $\omega=\Omega ( (8/3-n^2)/2)^{1/2}$.
  These constitute Eqs.~(10) and (11) in the main text.
\end{itemize}


\end{document}